\documentclass[twocolumn,showpacs,prl,amsmath,amssymb,superscriptaddress]{revtex4}
\usepackage{graphicx}% Include figure files
\usepackage{dcolumn}% Align table columns on decimal point
\usepackage{bm}% bold math
\usepackage{amssymb}

\begin{document}

\title{Huge nonequilibrium magnetoresistance in  hybrid superconducting spin valves}

\author{Francesco Giazotto}
\email{giazotto@sns.it}
\affiliation{NEST CNR-INFM and Scuola Normale Superiore, I-56126 Pisa, Italy}
\author{Fabio Taddei}
\affiliation{NEST CNR-INFM and Scuola Normale Superiore, I-56126 Pisa, Italy}
\author{Rosario Fazio}
\affiliation{NEST CNR-INFM and Scuola Normale Superiore, I-56126 Pisa, Italy}
\affiliation{International School for Advanced Studies (SISSA), I-34014 Trieste, Italy}
\author{Fabio Beltram}
\affiliation{NEST CNR-INFM and Scuola Normale Superiore, I-56126 Pisa, Italy}

%\date{\today}

\begin{abstract}
A hybrid ferromagnet-superconductor spin valve is proposed. Its operation relies on the interplay between nonequilibrium transport and proximity-induced exchange coupling in superconductors. Huge tunnel magnetoresistance values as large as some $10^6\%$ can be achieved in suitable ferromagnet-superconductor combinations under proper voltage biasing. The controllable spin-filter nature of the structure combined with its intrinsic simplicity make this setup attractive for low-temperature spintronic applications where reduced power dissipation is an additional requirement.     
\end{abstract}


\maketitle
   
During the last decade research on novel magnetoresistive effects in mesoscopic structures has gained increasing interest both for fundamental physics and in the context of \emph{spintronic} nanodevice applications \cite{fabian}. 
Much of this effort focused on the spin valve effect whose prototype is represented by the "giant magnetoresistance" (GMR) in magnetic multilayers \cite{baibich}.
GMR represents a resistance variation of the order of $100\%$ and is currently widely exploited in magnetic sensors and information storage applications.
More recently ferromagnet-superconductor systems came under the spotlight. In the latter magnetic order and superconductivity are combined to give rise
to exotic phenomena spanning from $\pi$ Josephson junctions \cite{buzdin} to Andreev reflection-induced magnetoresistance \cite{CAR}.

In this Letter we propose a hybrid ferromagnet-superconductor (FS) spin valve that can yield huge tunnel magnetoresistance values as high as several $10^6\%$. Operation is based on the interplay between out-of-equilibrium quasiparticle dynamics and proximity-induced exchange coupling in superconductors. This leads to a fully-tunable structure which shows high potential for application in spintronics.  

\begin{figure}[t!]
\includegraphics[width=\columnwidth,clip]{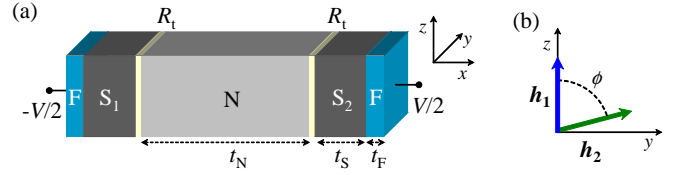}
\caption{(color) (a) Scheme of the FS-I-N-I-SF spin valve. Ferromagnetic layers (F) induce in each superconductor through proximity effect an exchange field whose relative orientation is controlled by an externally applied magnetic field. 
(b) The F exchange fields ($\boldsymbol{h_{1,2}}$) are confined to the $y-z$ plane, and are misaligned by an angle $\phi$.}
\label{fig1}
\end{figure}

The system we consider [see Fig. \ref{fig1}(a)] is a multilayer consisting of two identical FS bilayers (FS$_{1,2}$) symmetrically connected to a $t_{\text{N}}$-thick normal metal region (N) through insulating barriers (I) of resistance $R_{\text{t}}$. $t_{\text{F}}$ ($t_{\text{S}}$) labels the F (S) layer thickness and a voltage $V$ is applied across the structure. The exchange field in the left ferromagnet ($\boldsymbol{h_{1}}$) is aligned along the $z$ axis, while that in the right F layer ($\boldsymbol{h_{2}}$) is misaligned by an angle $\phi$ [see Fig. \ref{fig1}(b)]. For simplicity we set $|\boldsymbol{h_{1}}|=|\boldsymbol{h_{2}}|=h$.
Note that in one such structure $\boldsymbol{h_{2}}$ can be rotated by applying an in-plane magnetic field as low as some mT.
Finally we assume that (i) the FS interface is transparent \cite{prox} and (ii) $R_{\text{t}}$ is much larger than both the resistance of the N layer and the FS contact resistance (this ensures that each FS bilayer is in local equilibrium).
The electronic properties of a FS bilayer can be analyzed within the quasiclassical formalism \cite{buzdin}. We are interested in the situations in which the influence of the F layer on the superconductor becomes nonlocal. This occurs in the limit $t_{\text{S}}<\xi_{\text{S}}=\sqrt{ \hbar D/2\pi k_{\text{B}} T_{\text{c}}}$ and $t_{\text{F}}<\xi_{\text{F}}=\sqrt{ \hbar D/h}$, where $\xi_{\text{S}}$ and $\xi_{\text{F}}$ are the S coherence length and the length of the condensate penetration into the ferromagnet, respectively.
When this occurs, the ferromagnet induces in S a homogeneous \emph{effective} exchange field and modifies the superconducting gap ($\Delta$).
The effective values of the exchange field ($h^*$) and S gap ($\Delta^*$) are given by \cite{bergeret}: 
\begin{equation}
\begin{array}{l}
\Delta^*/\Delta=\nu_{\text{S}}t_{\text{S}}(\nu_{\text{S}}t_{\text{S}}+\nu_{\text{F}}t_{\text{F}})^{-1}\\
h^*/h=\nu_{\text{F}}t_{\text{F}}(\nu_{\text{S}}t_{\text{S}}+\nu_{\text{F}}t_{\text{F}})^{-1},
\end{array}
\end{equation}
where $\nu_{\text{S}}$ ($\nu_{\text{F}}$) is the normal-state density of states (DOS) in S (F).
If $\nu_{\text{F}}=\nu_{\text{S}}$ and  for $t_{\text{F}}\ll t_{\text{S}}$, it follows that $\Delta^*\simeq\Delta$ while $h^*/h\simeq t_{\text{F}}/t_{\text{S}}\ll1$.
We assume that the only effect of $h^*$ on the quasiparticles consists in leading to a spin-dependent superconducting DOS. Namely, the latter will be BCS-like, but shifted by the effective exchange energy (equivalent to that of a Zeeman-split superconductor in a magnetic field \cite{meservey}), $\mathcal{N}^{\text{S}}_{\sigma}(\varepsilon)=|\text{Re}[(\varepsilon+\sigma h^*+\Gamma)/2\sqrt{(\varepsilon+\sigma h^*+\Gamma)^2-\Delta^{*2}}]|$, where $\varepsilon$ is the energy measured from the condensate chemical potential, $\Gamma$ is a smearing parameter \cite{pekola,calculation}, and $\sigma =\pm 1$ refers to spin parallel (antiparallel) to $z$.
\begin{figure}[t!]
\includegraphics[width=\columnwidth,clip]{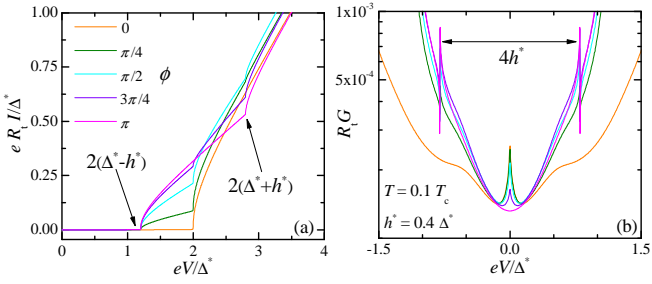}
\caption{(color) (a) Current vs bias voltage $V$ for several angles $\phi$ at $T=0.1T_{\text{c}}$ and $h^*=0.4\Delta^*$. (b) Differential conductance $G$ vs $V$ calculated for the same values as in (a). 
}
\label{tS}
\end{figure}

At finite bias $V$ and in the limit of negligible inelastic scattering, the steady-state nonequilibrium distribution functions in the N layer are spin dependent and can be written as
\begin{equation}
f_{\sigma}(\varepsilon,V,h^*,\phi)=\frac{\mathcal{N}^{\text{S}_1}_{\sigma}\mathcal{F}^{\text{S}_1}+[a(\phi)\mathcal{N}^{\text{S}_2}_{\sigma}+b(\phi)\mathcal{N}^{\text{S}_2}_{-\sigma}]\mathcal{F}^{\text{S}_2}}{\mathcal{N}^{\text{S}_1}_{\sigma}+a(\phi)\mathcal{N}^{\text{S}_2}_{\sigma}+b(\phi)\mathcal{N}^{\text{S}_2}_{-\sigma}},
\label{distribution}
\end{equation}
where $a(\phi)=\text{cos}^2[\phi/2]$, $b(\phi)=\text{sin}^2[\phi/2]$, $\mathcal{F}^{\text{S}_1(\text{S}_2)}=f_0(\varepsilon\pm eV/2)$,
$\mathcal{N}^{\text{S}_1}_{\sigma}=\mathcal{N}^{\text{S}}_{\sigma}(\varepsilon +eV/2)$, $\mathcal{N}^{\text{S}_2}_{\sigma}=\mathcal{N}^{\text{S}}_{\sigma}(\varepsilon -eV/2)$, $f_0(\varepsilon)$ is the Fermi function at temperature $T$, and $e$ is the electron charge. 
Equation (2) is obtained by generalizing the results of Ref. \cite{giazotto} to noncollinear exchange fields \cite{relaxation}.
The electronic transport properties of the structure are entirely determined by the spin-dependent distribution functions in (\ref{distribution}).
In particular, the quasiparticle current $I$ (evaluated in the left interface) is given by $I=\sum_{\sigma}I_{\sigma}$, where
\begin{equation}
I_{\sigma}(V,h^*,\phi)=\frac{1}{eR_{\text{t}}}\int^{\infty}_{-\infty}d\varepsilon \mathcal{N}^{\text{S}_1}_{\sigma}(\varepsilon)[\mathcal{F}^{\text{S}_1}(\varepsilon)-f_{\sigma}(\varepsilon)].
\end{equation}

Figure 2(a) displays the electric current vs bias voltage $V$ calculated for several angles $\phi$ at $h^*=0.4\Delta^*$ and $T=0.1T_{\text{c}}$, where $T_{\text{c}}=(1.764k_{\text{B}})^{-1}\Delta^*$.  
This figure can be understood in terms of the energy shift of the spin-dependent DOS of S$_{1,2}$ caused by the applied $V$ and $h^*$.
A sizable current starts to flow only when the voltage $V$ is such that the DOS is finite for both S in some range of energies.
For $\phi=0$, the current rises sharply at $|eV|=2\Delta^*$, similarly to the quasiparticle current of a SIS junction (also in the presence of an in-plane magnetic field \cite{meservey}). 
In this case, in fact, the DOS of a given spin is shifted by the Zeeman energy in the same direction for both superconductors.
In contrast, for $\phi=\pi$ current sets off at $|eV|=2(\Delta^*- h^*)$. In this case the DOS of S$_{1}$ is shifted in the opposite direction with respect to that of S$_2$ and the required voltage for a current to flow is smaller with respect to the $\phi=0$ case.
It is easy to understand the origin of additional feature appearing at $|eV|=2(\Delta^*+ h^*)$.
For intermediate values of $\phi$, features are present at $|eV|=2(\Delta^*\pm h^*)$ and at $|eV|=2\Delta^*$. These stem from contributions linked to both $\phi=0$ and $\phi=\pi$ configurations.
Of particular relevance is the voltage interval $2(\Delta^*-h^*)\leq |eV|\leq 2\Delta^*$. By increasing $\phi$ from $0$ to $\pi$, the current is enhanced from a vanishingly small value up to a finite value leading to a spin-valve effect.
Figure 2(b) shows the differential conductance $G(V)=dI/dV$  calculated for the same values as in Fig. 2(a). Additional features are present at $|eV|=2h^*$ which are strongly temperature-dependent, and vanish in the limit $T \rightarrow 0$ \cite{features} (the zero-bias conductance peak for $\phi\neq\pi$ resembles that typical of a SIS junction composed of identical superconductors \cite{tinkham}).
It is noteworthy to mention that the nonequilibrium condition is essential for the observation of the spin-valve effect. At equilibrium the distribution functions in the N layer would be thermal and spin-independent.
\begin{figure}[t!]
\includegraphics[width=\columnwidth,clip]{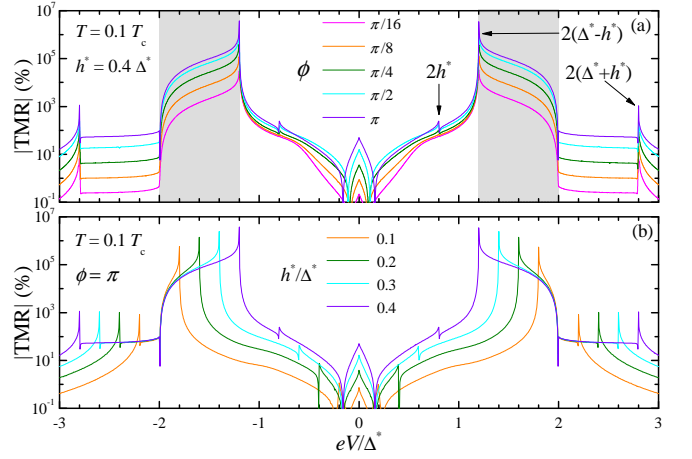}
\caption{(color) (a) Nonequilibrium tunnel magnetoresistance ratio $|\text{TMR}|$ vs $V$ calculated for several angles $\phi$ at $T=0.1T_{\text{c}}$ and $h^*=0.4\Delta^*$. Gray regions correspond to voltage intervals of very large TMR. (b) $|\text{TMR}|$ vs $V$  for different $h^*$ values at $T=0.1T_{\text{c}}$ and $\phi=\pi$.}
\label{tF}
\end{figure}

The spin-valve properties of the present FS-I-N-I-SF setup can be evaluated quantitatively by analyzing the tunnel magnetoresistance ratio ($\text{TMR}$), defined as
\begin{equation}
\text{TMR}(V,h^*,\phi)=\frac{G(V,h^*,\phi)-G(V,h^*,0)}{G(V,h^*,0)}.
\end{equation}
Figure 3(a) displays the absolute value of the TMR vs bias voltage $V$ calculated for several angles $\phi$ at $T=0.1T_{\text{c}}$ and $h^*=0.4\Delta^*$. For clarity, the regions corresponding to large TMR are highlighted by gray shading. 
For $2(\Delta^*-h^*)\leq |eV|\leq 2\Delta^*$ the TMR increases monotonically by increasing $\phi$ and is maximized at $\phi=\pi$ where it obtains huge values larger than $10^6\%$. We note that in the limit $T=0$ and $\Gamma=0$ $|\text{TMR}| \rightarrow \infty$, i.e., ideal full spin-valve effect.  
The TMR behavior for several  exchange field values is shown in Fig. 3(b), at $T=0.1T_{\text{c}}$ and $\phi=\pi$. By decreasing $h^*$, the maximum TMR value reduces, and so does the voltage interval of larger magnetoresistance.  Larger $h^*$ values are thus preferable in order to extend the voltage window for optimized operation and to maximize TMR values. 

The experimental accessibility of these effects must be carefully assessed. Up to this point we focused on the $full$ nonequilibrium regime. TMR values, however, can be expected to be marginally affected also in the presence of electron-electron relaxation in the N layer.
In fact Coulomb interaction allows quasiparticles to exchange energy (through inelastic collisions) without coupling the two spin species.
This was verified by solving a kinetic equation which takes into account screened Coulomb interactions \cite{relaxation}.
TMR is virtually unmodified even in the perfect \emph{quasiequilibrium}  limit \cite{giazotto,RMP} (i.e., where  $f_{\sigma}$ functions are forced into thermal distributions at a temperature different from that of the lattice).
By contrast, TMR decreases if spin-flip processes mix the spin-dependent distributions. In metals and at low temperature (below $\sim1$ K), such processes are caused by the presence of magnetic impurities in the N layer. 
Spin-flip scattering can be suppressed by limiting the magnetic-impurity content in the N layer, and by choosing $t_{\text{N}}\ll \lambda_{\text{sf}}$, where
the spin-flip relaxation length $\lambda_{\text{sf}} \sim \mu$m in metals such as Cu or Au \cite{jedema,johnson}). These constraints can be met fairly easily experimentally in a multilayered structure like the one presented here.  

The spin-filtering properties of our device can be quantified by inspecting the current polarization ($P_I$), defined as
\begin{equation}
P_I(V,h^*,\phi)=\frac{I_{+}(V,h^*,\phi)-I_{-}(V,h^*,\phi)}{I_{+}(V,h^*,\phi)+I_{-}(V,h^*,\phi)}.
\end{equation}
The calculated nonequilibrium $P_I$ vs $V$ is displayed in Fig. 4 (a) for several $\phi$ values, at $T=0.1T_{\text{c}}$ and $h^*=0.4\Delta^*$. Upon increasing $\phi$, two intervals of $100\%$ spin-polarized current develop for $2(\Delta^*-h^*)\leq |eV|\leq 2\Delta^*$, extending to wider regions [$2(\Delta^*-h^*)\leq |eV|\leq 2(\Delta^*+h^*)$] as $\phi$ approaches $\pi$ \cite{nota}.
Depending on the bias, fully spin-polarized currents of both parallel and antiparallel spin species can be obtained \cite{polarization}. The structure can thus be also operated as a \emph{controllable} spin-filter by changing the orientation of $\boldsymbol{h_{2}}$ as well as by varying $V$. Figure 4(b) shows $P_I$ vs $V$ for several $h^*$ at $T=0.1T_{\text{c}}$ and $\phi=\pi$. The net effect of increasing $h^*$  is to widen the regions of $100\%$ spin-polarized current.
\begin{figure}[t!]
\includegraphics[width=\columnwidth,clip]{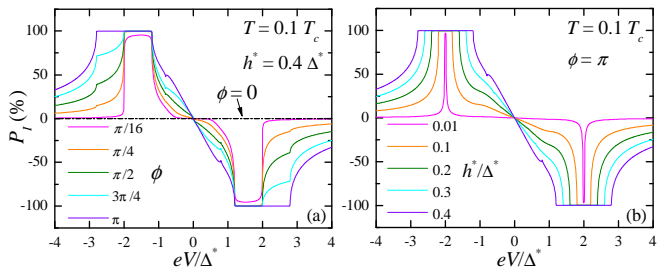}
\caption{ (color) (a) Nonequilibrium current polarization $P_I$ vs $V$ calculated for several angles $\phi$ at $T=0.1T_{\text{c}}$ and $h^*=0.4\Delta^*$. (b) $P_I$ vs $V$ for different $h^*$ values at $T=0.1T_{\text{c}}$ and $\phi=\pi$.}
\label{exch}
\end{figure}

A first application of the present structure is the implementation of storage cell elements, thanks to the very large  TMR values [see Fig. 3(a)]. Magnetic-field-controlled current switches can also be envisioned [see Fig. 2(a)].
Importantly, power dissipation is intrinsically very limited owing to the small currents driven through  SIN junctions. For example, assuming $R_{\text{t}}=10^3\,\Omega$ and aluminum (Al) electrodes at $T=0.1T_{\text{c}}\approx  0.12$ K, a dissipated power in the range $\sim10^{-15}-10^{-12}$ W can be achieved for $2(\Delta^*-h^*)/e<|V|<2\Delta^*/e$. This makes this setup attractive for low-dissipation cryogenic applications. We note that although similar results could be obtained in a FS-I-SF structure (i.e., without the N interlayer) and not relying on nonequilibrium, the present system possesses a crucial advantage. 
In fact a FS-I-SF structure implies an additional undesired Josephson current. The latter could be suppressed, for example, by the application of an additional in-plane magnetic field. This field, however, would largely exceed that required to control the orientation of $h^*$.

In light of a realistic implementation, ferromagnetic alloys such as Cu$_{1-x}$Ni$_x$ \cite{ryazanov} or Pd$_{1-x}$Ni$_x$ \cite{kontos} (which allow fine tuning of $h$ through a proper choice of $x$) are promising candidates. For example, in Pd$_{1-x}$Ni$_x$ alloy with $x=0.1$, $h\simeq 10$ meV resulting in $\xi_{\text{F}}\approx 5$ nm \cite{kontos}. By choosing Al as S electrodes (with $\Delta\simeq 200\,\mu$eV and $\xi_{\text{S}}\approx 300$ nm \cite{romijn}) it turns out that $h^*$ in the range $\sim0.2\Delta^*-0.5\Delta^*$ can be achieved. 

In conclusion, we have proposed a hybrid ferromagnet-superconductor tunable spin valve which allows very large tunnel magnetoresistance values ($\text{TMR}\sim 10^6\%$). Its operation relies on the interplay between  nonequilibrium and proximity-induced exchange coupling in superconductors. The simplicity intrinsic to the structure combined with limited power dissipation makes this setup promising for the implementation of low-temperature magnetoresistive devices.   
Partial financial support from MIUR under FIRB program RBNE01FSWY is acknowledged.


\begin{thebibliography}{99}
\bibitem{prox}
	In the opposite case, the superconductor is marginally affected by the proximity of the F layer \cite{buzdin}.
\bibitem{calculation}
	We set throughout the paper a realistic smearing parameter $\Gamma=10^{-4}\Delta^*$ (see Ref. \cite{pekola}).
\bibitem{relaxation}
	Further details will be reported elsewhere.
\bibitem{features}
	We stress once more that all the finite-bias features appearing in Figs. 2(a,b) simply reflect how the spin-dependent DOS in each superconductor contributes to the total quasiparticle current at different $V$.
\bibitem{polarization}
	Note that $P_I=100\%$ is achievable in the quasiequilibrium limit as well.
\end{thebibliography}
\end{document}